\documentclass[10pt,letterpaper,aps,prl,twocolumn,superscriptaddress,floatfix,footinbib]{revtex4-1}
\usepackage[latin1]{inputenc}
\usepackage{amsmath}
\usepackage{amsfonts}
\usepackage{amssymb}
\usepackage{bm}
\usepackage[pdftex]{graphicx}
\usepackage[colorinlistoftodos]{todonotes}  
\usepackage{graphicx}
\usepackage{subfigure}    
\newcommand{\ket}[1]{\lvert#1\rangle}

\newcommand{\la}{\langle}
\newcommand{\ra}{\rangle}
\newcommand{\be}{\begin{equation}}
\newcommand{\ee}{\end{equation}}
\newcommand{\Tr}{\rm Tr}

\begin{document}
\title{Arrow of Time for Continuous Quantum Measurement}
\author{Justin Dressel}
\affiliation{Institute for Quantum Studies, Chapman University, Orange, CA 92866, USA}
\affiliation{Schmid College of Science and Technology, Chapman University, Orange, CA 92866, USA}
\author{Areeya Chantasri}
\affiliation{Department of Physics and Astronomy, University of Rochester, Rochester, NY 14627, USA}
\affiliation{Center for Coherence and Quantum Optics, University of Rochester, Rochester, NY 14627, USA}
\author{Andrew N. Jordan}
\affiliation{Department of Physics and Astronomy, University of Rochester, Rochester, NY 14627, USA}
\affiliation{Center for Coherence and Quantum Optics, University of Rochester, Rochester, NY 14627, USA}
\affiliation{Institute for Quantum Studies, Chapman University, Orange, CA 92866, USA}
\author{Alexander N. Korotkov}
\affiliation{Department of Electrical and Computer Engineering, University of California, Riverside, CA 92521, USA}
\date{\today}
\begin{abstract}
We investigate the statistical arrow of time for a quantum system being monitored by a sequence of measurements. For a continuous qubit measurement example, we demonstrate that time-reversed evolution is always physically possible, provided that the measurement record is also negated. Despite this restoration of dynamical reversibility, a statistical arrow of time emerges, and may be quantified by the log-likelihood difference between forward and backward propagation hypotheses. We then show that such reversibility is a universal feature of non-projective measurements, with forward or backward Janus measurement sequences that are time-reversed inverses of each other. 
\end{abstract}
\maketitle

The classical dynamics of a conservative system is time-reversible. If we watch a movie backwards in the absence of friction, it will show dynamics perfectly consistent with the laws of motion, so we may not distinguish whether we watch the movie forward or backwards in time from the dynamics alone. However, when the system has more than a few degrees of freedom---such as during the starting break in a game of pool---then the likelihood that the evolution is either forward or backward in time may differ, so it becomes possible to distinguish an arrow of time statistically. The existence of such an arrow of time is a fundamental question, and has been of interest in many areas of physics \cite{Hawking1985,Lebowitz1993,hartle2014quantum}.   

The quantum dynamics of a conservative and unmeasured system is similarly time-reversible. For example, the Schr{\"o}dinger equation becomes invariant under time-inversion if the position-space wavefunction is complex-conjugated. This is a special case of a general anti-unitary time-reversal operation \cite{gottfried2013quantum}, and is sufficient to restore time symmetry for a closed quantum system. 

The introduction of a sequence of measurements seems to break such dynamical symmetry, however, for two distinct reasons. First, obtaining definite measurement results traditionally collapses the wavefunction, which produces non-unitary evolution that is distinct from the Schr{\"o}dinger equation and not reversed by the same anti-unitary operation. Second, the randomness of each measurement creates an intrinsic asymmetry between an unknown future and a definite past. These reasons have contributed to the view that quantum mechanics is fundamentally asymmetric in time \cite{watanabe1965conditional,Smolin2014}.

We seek to clarify this apparent discrepancy between classical and quantum reversibility. In the past, such efforts to restore reversibility have tried reformulating quantum mechanics in a more symmetric way \cite{gell1994time}. For example, the ``two-time'' formalism of Aharonov, Bergmann, and Lebowitz \cite{Aharonov1964time} removes the indefiniteness of the future by introducing a second boundary condition (or postselection) that brackets a time interval, and avoids non-unitary state collapse by considering infinitesimally weak measurements that do not affect the state within that interval \cite{Aharonov1988}. Physical measurements have nonzero strength, however, so will (at least partially) collapse the state and seemingly spoil the reversibility of such a scheme \cite{Maccone2009}. Nevertheless, partial collapses of the state may still be fully restored probabilistically (``wavefunction uncollapse''), even if the initial state is unknown \cite{Korotkov2006undo,Andrew2010uncollapse}. This uncollapsing phenomenon has been confirmed experimentally in superconducting and optical systems \cite{Korotkovexp2008,Kim2009,Zhong2014}, which raises the question once more whether the time symmetry of a sequence of several such measurements could be similarly restored. 

\begin{figure}[t]
\includegraphics[width=0.9\columnwidth]{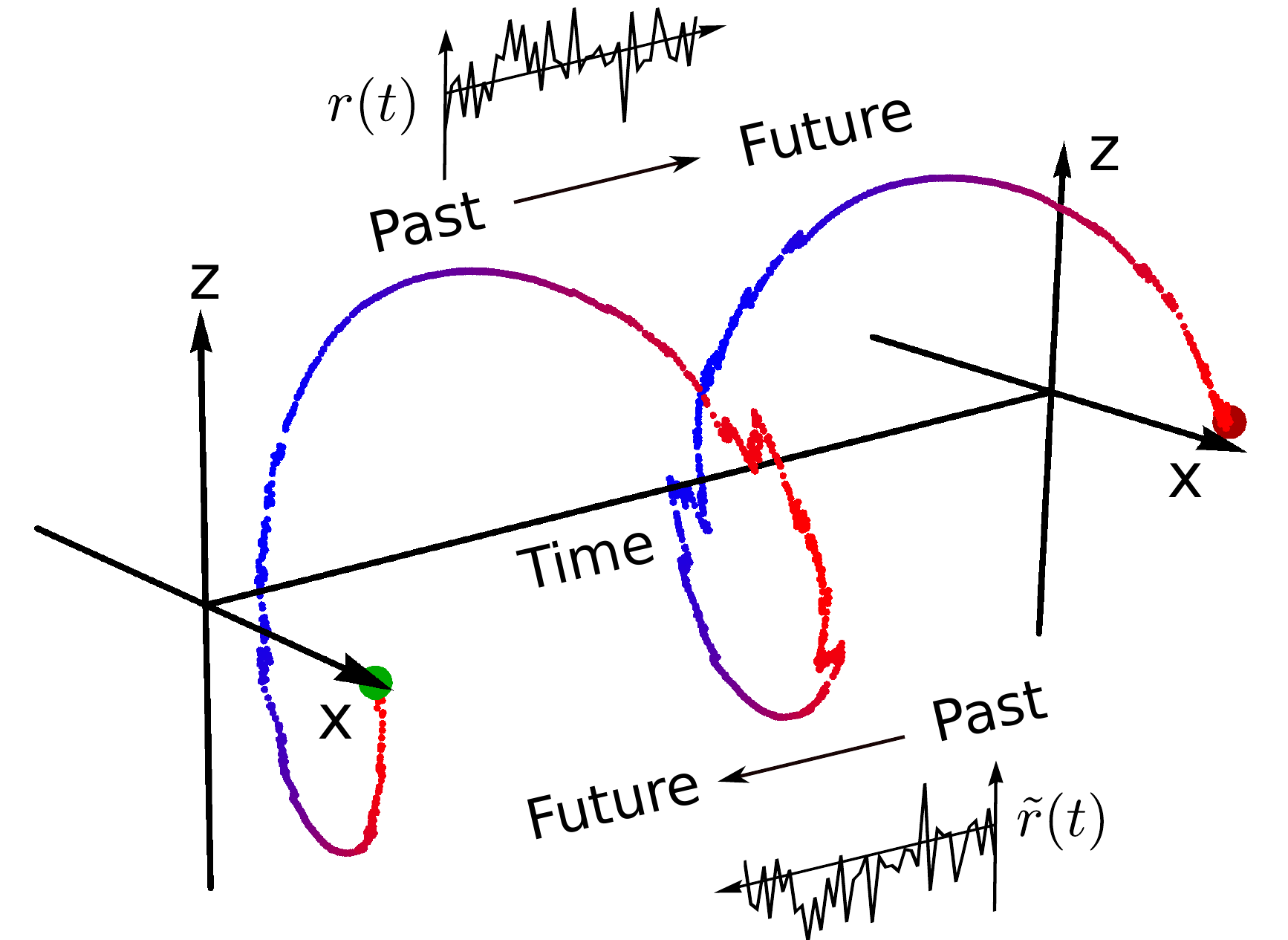}
\caption{A single quantum trajectory of a continuously monitored qubit.  The $x$ and $z$ Bloch sphere coordinates of a qubit change due to both unitary and measurement dynamics. The red and blue colors denote positive and negative values of the $x$ coordinate. The boundary states are shown as green and red dots. Is time running forward with measurement record $r(t)$ or backward with flipped record $\tilde{r}(t)$?}
\label{helical}
\end{figure}

In this Letter, we demonstrate how to restore time reversal symmetry for a sequence of nonprojective measurements that takes into account the insights from measurement uncollapse. This is a nontrivial problem, since correlation functions of even arbitrarily weak, ostensibly non-invasive, measurements break time-reversal symmetry in general \cite{bednorz2013noninvasiveness,Diosi2016}. 
We solve the general problem by considering two complementary measurement sequences, one pointing into the future, and another into the past, that are time-reversed inverses of each other. We name these complements \emph{Janus sequences}. For qubits, this general solution takes a particularly simple form that can be taken to the limit of time-continuous measurements, 
producing so-called quantum trajectories \cite{BookCarmichael,BookWallMilburn,Wiseman1993,Korotkov1999,Goan2001,Korotkov2001}. Quantum trajectory theory eliminates any remaining separation between Schr{\"o}dinger equation dynamics and measurement disturbance, and replaces them with a single stochastic process that includes both. 

We pose the time-reversibility problem in the following way: 
Suppose we are given a movie of stochastic quantum state dynamics along with its associated noisy detector output (a sort of ``soundtrack'' for the movie). We are then asked to determine whether the movie shows the forward evolution of the state, or whether the movie has been reversed, as depicted in Fig.~1. In the simplest case of a monitored qubit, we find that such a movie played backwards obeys time-reversed equations of motion if we also flip the sign of its soundtrack (measurement record). We stress that this is {\it not} a microscopic time-reversal of the measurement apparatus, nor is it a backward inference (past quantum state) kind of dynamics \cite{Gammelmark2013,Dressel2013}---our time reversal shows equally valid forward dynamics.
After watching the movie for a longer duration while listening to its soundtrack, we can distinguish a forward from a time-reversed movie with increasing certainty in order to probabilistically find the arrow of time.  This phenomenon of time-symmetry with a time-arrow is closely analogous to classical physics, with an important difference that in classical physics the initial state is typically special, whereas in the quantum measurement case it is the final state that is typically special.

Note that to achieve perfect time reversal we must not lose information to the environment, other than to an ideal, quantum-limited, detector. That is, we must consider a system that is being monitored without additional noise or ``quantum friction'', just as in classical physics. As a physical example, a superconducting qubit like a transmon \cite{koch2007charge} may be continuously monitored with microwaves using circuit quantum electrodynamics \cite{murch2013observing,weber2014mapping}, yielding a time-dependent (noisy) homodyne quadrature readout $I(t)$; such monitoring would yield time-reversible evolution if the amplifier were quantum limited with no loss in the readout chain (i.e., no readout inefficiency) \cite{hatridge2011dispersive}, and were otherwise decoupled from its environment. 

Similarly, the current $I(t)$ flowing through a quantum point contact can continuously monitor a double-quantum dot charge qubit with nearly quantum-limited efficiency \cite{Korotkov1999,Pilgram2002,Clerk2003}.

{\it Continuous qubit measurement.}---%
We consider such a quantum-limited continuous qubit measurement as an illustrative example, which we will later generalize to an arbitrary sequence of measurements.  
Specifically, continuously monitoring the $\sigma_z$ observable produces a noisy informational signal of the form $I(t) = \bar{I} + (\Delta I/2) z(t) + \sqrt{S/2}\,\xi(t)$,
where $\Delta I \equiv I_1 - I_0$ is the difference between the average signals observed for definite qubit $\ket{1}$ and $\ket{0}$ states, $\bar{I}$ is the average background signal, $z(t) = \Tr{[\sigma_{z} \rho(t)]}$ is the Bloch sphere $z$-component of the qubit state, and $\sqrt{S/2}\,\xi(t)$ is a (white)  Gaussian noise process, $\langle \xi(t)\xi(0)\rangle = \delta(t)$, with a spectral density $S$ that arises from, e.g., quantum vacuum fluctuations \cite{murch2013observing}. After rescaling the signal as $r(t) \equiv 2(I(t)-\bar{I})/\Delta I$ and defining the characteristic {\it measurement time} (or inverse measurement strength) $\tau \equiv 2S/(\Delta I)^2$ for achieving unit signal-to-noise ratio, we find $r(t) = z(t) + \sqrt{\tau}\,\xi(t)$.
We also assume a qubit Hamiltonian, given by $H = \hbar \Omega \sigma_y/2$, produced (for example) by a microwave drive, which causes rotation in the $x$-$z$ plane of the Bloch sphere. 

The resulting quantum trajectory equations for the qubit state are 
\begin{align}\label{qbayes}
{\dot x} = -\Omega z - \frac{x z r}{\tau}, \;\;  
{\dot y} = - \frac{y z r}{\tau} ,  \;\;
{\dot z} = \Omega x + \frac{(1-z^2) r}{\tau}, 
\end{align}
given in the (time-symmetric) Stratonovich picture \cite{Korotkov1999}, omitting explicit $t$-dependence. These equations assume ideal conditions, including efficient detection and Markovian evolution, so any residual entanglement between the qubit and detection apparatus is assumed to vanish (e.g., a microwave resonator must operate in the ``bad cavity'' limit). 
We observe that Eqs.~(\ref{qbayes}) are time reversal invariant under the transformations: $t \mapsto - t$,  $\Omega \mapsto -\Omega$, keeping $x,y,z$ invariant, provided that the \emph{record is also flipped}, $r \mapsto -r$. With these changes, a quantum movie for a single measurement run is the same when played backward, as illustrated in Fig.~1 for the special case of $y=0$ \footnote{We note that the Stratonovich formulation of the stochastic differential equations, with time-symmetric derivative, is advantageous for understanding the time-reversal symmetry of the evolution: this feature is not apparent in the It{\^o} formulation.}, thus restoring time symmetry.

{\it Arrow of time.}---%
We will now show that although such continuous qubit measurement dynamics is time reversal invariant, we can nevertheless probabilistically distinguish forward and backward evolution, yielding a statistical arrow of time. This task can be phrased as a hypothesis testing problem: 
is the movie shown in Fig.~1 of duration $T$ running forward (F) or backward (B)?  

To test these hypotheses, let the prior probabilities $P(F)$ and $P(B)=1-P(F)$ indicate our initial guess whether the movie is running forward or backward. Let 
$P_F(r(t)) = P(r(t)|\rho_i)$ be the probability density of obtaining the measurement record $r(t)$, supposing the movie is running forward from an initial state $\rho_i$; similarly, let $P_B(r(t)) = P(-r(T-t)|\rho_f)$ be the probability density that supposes the movie is running backward from a final state $\rho_f$. This last situation is equivalent to a forward trajectory with outcome $-r(T-t)$, starting from an ``initial'' state $\rho_f$.  
We then use Bayes' rule to compute the likelihood that the movie is running in the forward direction given the movie and its soundtrack,
\be
P(F| r(t)) = \frac{P_F(r(t)) P(F)}{P_F(r(t)) P(F)+P_B(r(t)) P(B)}.
\ee
If we have no {\it a priori} bias about this question, we set $P(B) = P(F) = 1/2$, to find the likelihood
\be
P(F| r(t)) = \frac{{\cal R}}{1+ {\cal R}}, \quad {\cal R} = \frac{P_F(r(t))}{P_B(r(t))}. \label{R}
\ee
We therefore conclude that we can make no statistical inference \emph{only} if the forward and backward probability densities are identical (i.e., the probability ratio ${\cal R} =1$). The logarithm of this ratio, $\ln \mathcal{R}$, is thus a natural discriminator, with positive values inferring forward motion and negative values inferring backward motion. 
The mean value $\overline{\ln \mathcal{R}}$ over forward-generated trajectories thus gives an estimate of the statistical arrow of time for continuous quantum measurement, also named the ``length of time's arrow'' \cite{feng2008length}.
It is similar to the relative entropy (also known as the Kullback-Leibler divergence) between forward and backwards distributions.  Researchers in nonequilibrium statistical physics have used analogous arrow-of-time hypothesis discrimination to quantify the entropy production (or irreversibility) of mesoscopic systems  \cite{kawai2007dissipation,jarzynski2011equalities,feng2008length,diosi2004probability,batalhao2015irreversibility}.  There has been recent cross-pollination of the methodology in these fields \cite{alonso2016thermodynamics,elouard2015stochastic}.
 
To find the relative probability densities of the trajectories $r(t)$ versus $-r(T-t)$, given a quantum trajectory, we may expand the distribution of results to first order in a small time-step to find
$P(r(t)|\rho_i) \propto \exp[ - \int_0^T dt' (r(t')^2 -2 r(t') z(t') +1)/2\tau]$ \cite{chantasri2013action,chantasri2015stochastic}, where the backwards distribution simply time-reverses the integral, and 
flips the sign of $r$ at every time \footnote{We note that the time integrals in this paper assume the Stratonovich interpretation, which uses a time-symmetric midpoint formulation for the Riemann sum.}.
The arrow of time ratio ${\cal R}$ Eq.~(\ref{R}) is given in terms of the probability densities of the forward trajectories $r(t)$ and the backward trajectories $-r(T-t)$, so \cite{Note2}
\be 
\ln {\cal R} = \frac{2}{\tau} \int_0^T dt\, r(t) z(t).  \label{calr}
\ee
This relative log-likelihood will then categorize each run of the experiment as being more likely to be running forward in time, $\ln {\cal R} >0$, or backward in time, $\ln {\cal R} <0$. In the latter case, we interpret Eq.~(\ref{calr}) to mean that the result $r(t)$ ``disagrees'' with the state component $z(t)$ it is estimating (has the opposite sign) more often than it ``agrees'' with it during the run, making reversed time-evolution more likely.
The average of this relative log-likelihood
may be simplified by writing the measurement result as $r(t) = z(t) + \sqrt{\tau} \xi(t)$, 
giving $\overline{\ln {\cal R}} = (1/\tau) \int_0^T dt \ (1+\la z(t)^2 \ra)$ after the stochastic average.

\begin{figure}[t]
\includegraphics[width=\columnwidth]{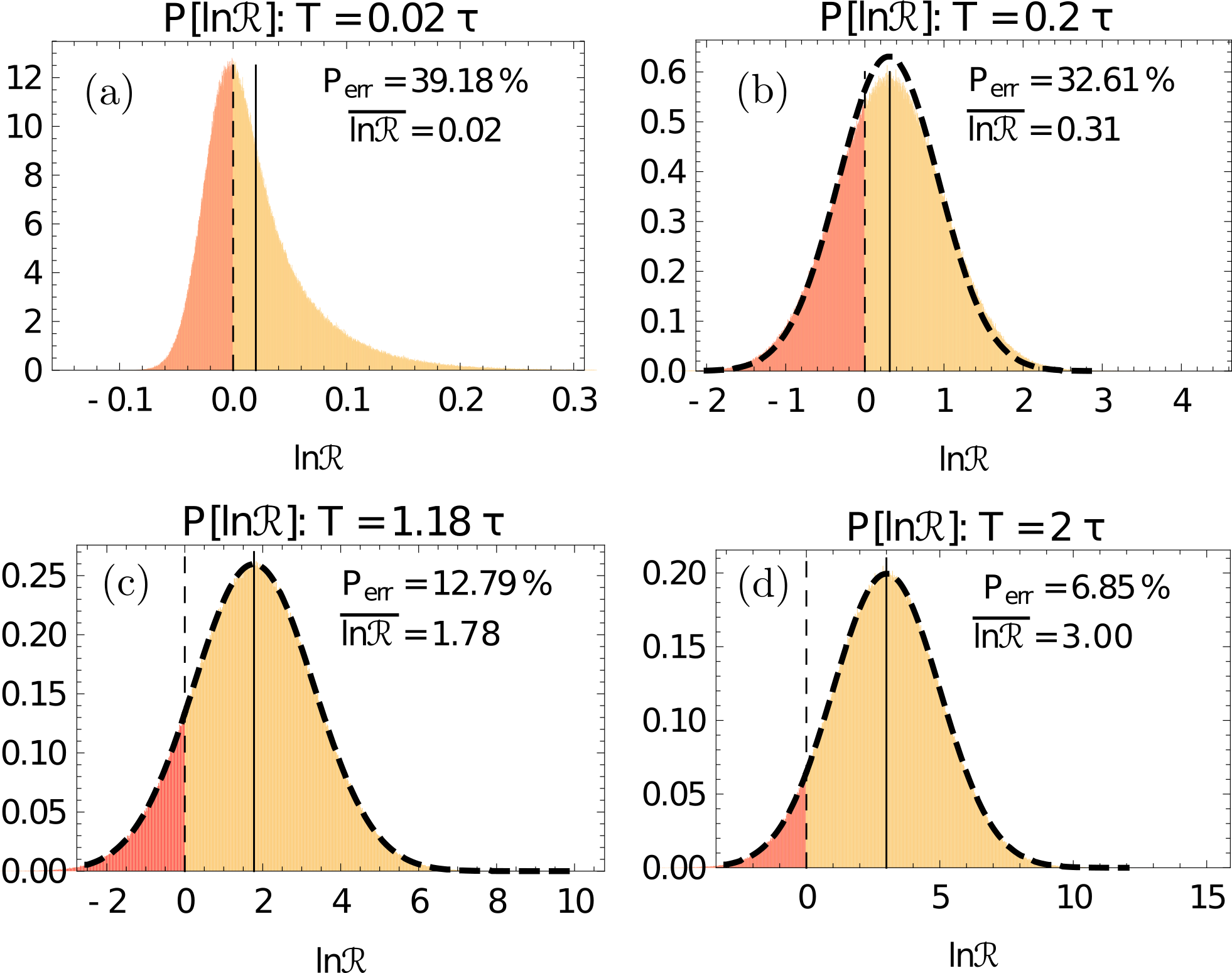}
\caption{Histograms of $\ln {\cal R}$ for $2\times10^6$ runs of monitored Rabi oscillations with period $2\pi/\Omega = 0.5\tau$ and measurement time $\tau$, starting from $x(t=0)=1$. The ratio of forward to backward probability distributions ${\cal R}$ discriminates the likelihood for a trajectory to be forward ($\ln {\cal R} > 0$) or backward ($\ln {\cal R} < 0$). The probability $P_{\rm err}$ of erroneously guessing reversed time is the red shaded area to the left of the vertical dashed line at $\ln{\cal R} = 0$. The vertical solid line is the mean $\overline{\ln {\cal R}}$, a measure of the statistical arrow of time. The durations $T = (0.02, 0.2, 1.18, 2.0) \tau$ are shown in subfigures (a--d), showing that for $T \geq 2\pi/\Omega$ (c,d), the distribution has converged to a Gaussian with mean $3T/2\tau$ and variance $2T/\tau$ (dashed profile in b,c,d).}
\label{histogram}
\end{figure}

{\it Numerical results for the arrow of time.}---%
Consider the case of persistent, diffusive, Rabi oscillations \cite{KorotkovAverin2001}, when $\Omega \gg \tau^{-1}$, so that the qubit performs oscillations in the $x$-$z$ plane with phase diffusion. 
For $T > 2\pi/\Omega$, the Rabi oscillations average to $\la z^2 \ra \approx \la (\cos \Omega t)^2 \ra = 1/2$, so
\be
\overline{\ln {\cal R}} \approx \frac{3T}{2\tau}. \label{final}
\ee
In this case, the average distinguishability of the forward from the backward arrow of time increases linearly with the duration of the measurement run.  

This statement may be made more precise by examining the entire distribution of $\ln {\cal R}$, which is shown from numerical simulations in Fig.~2 using a Rabi period of $2\pi/\Omega = 0.5\tau$. For durations longer than the Rabi period, $T > 2\pi/\Omega$, the mean grows linearly with the duration of the experiment, as predicted in (\ref{final}).   A calculation similar to the mean gives an approximate variance of $2T/\tau$.  The full distribution of $\ln {\cal R}$ becomes a broad Gaussian in this regime of $T > 2\pi/\Omega$, with the aforementioned mean and variance, see Fig.~2c,d.
Thus, the probability of erroneously guessing a forward trajectory to be backward is the area of the negative tail of the Gaussian 
\be 
P_{\rm err} \approx \frac{1}{2}\left[1 - {\rm{Erf}}\left(\frac{3}{4}\sqrt{\frac{T}{\tau}}\right)\right].
\ee
Consequently, if we wish to distinguish the forward-in-time arrow from the backward arrow to greater than $n$ standard deviations, we require a duration $T \ge 8 n^2 \tau/9$.  As can be seen from the histograms in Fig.~2, even for reasonably long duration $T$ it is common to observe readouts that appear to be reversed. We show examples of this, as well as present the simpler no-drive case, in the Supplementary Material. 

{\it Janus measurement sequences.}---%
We now generalize the simple qubit example to arbitrary sequences of generalized measurements. First, we reverse the direction of time for Schr\"odinger equation evolution in the standard way \cite{sakurai2011modern}, by introducing an anti-unitary time-reversal operation $\Theta$, satisfying $\la \Theta \Phi | \Theta \Psi \ra = \la \Psi | \Phi \ra$. In the case of position wavefunctions, $\Theta$ is simply the complex conjugate operation. More generally, $\Theta$ must correctly time-reverse all physical observables such as position, $\Theta {\bf x} \Theta^{-1} = {\bf x}$, momentum $\Theta {\bf p} \Theta^{-1} = - {\bf p}$, and spin $\Theta {\bf S} \Theta^{-1}  = -{\bf S}$, 
as well as the sign of any external magnetic field, $B\to-B$. Applying the time-reversal operator $\Theta$ to a quantum state $|\Psi(t)\ra$ inverts its temporal meaning, such that forward unitary time evolution $U_t$ correctly rewinds the dynamics: $U_t \Theta |\Psi(t)\ra = \Theta |\Psi(0)\ra$. 

Second, we add sequences of generalized measurements to the unitary dynamics. We first consider a forward sequence of measurements in time, $A, B, C, \ldots$, which will be one of two distinct Janus sequences that we will need. This sequence has possible measurement results $j = a, b, c, \ldots$, each of which will partially collapse the quantum state according to a measurement operator, $M_a, M_b, M_c, \ldots$. An initial state $|\Psi\ra$ thus evolves into 
\be
 | \Phi \ra  \propto       \ldots  M_c  M_b  M_a |\Psi\ra \equiv M_F |\Psi\ra,
\ee
where $M_F \equiv \ldots M_c M_b M_a$. Note that we include any intermediate unitary time evolution $U_t$ inside the Kraus operators. This formulation is quite general, so the measurement results may be discrete or continuous variables.   

We next introduce a corresponding backwards Janus sequence, which is a series of (in general different) measurements $A', B', C', \ldots$, with outcomes $j' = a', b', c', \ldots$ and Kraus operators $M_B\equiv M_{a'}M_{b'}M_{c'}\ldots$ also applied sequentially to the system, but in reverse order to the time-reversed ``initial'' state $\Theta |\Phi \ra$.  
Crucially, for {\it some possible} results $(j, j')$ of both sequences, we wish for the system state to rewind its path, restoring the initial (time-reversed) state: $M_B \Theta |\Phi\ra \propto \Theta |\Psi \ra$.
We can find the condition for this to happen by inserting ${\bf 1} = \Theta^{-1} \Theta$ between every pair of operators, yielding 
\be
M_{j'} \propto (\Theta M_j \Theta^{-1})^{-1}, \quad (j,j') = (a,a'),(b,b'),\ldots \label{janus}
\ee
That is, each measurement operator of the backward Janus sequence must be proportional to the {\it inverse} time-reversed measurement operator of the forward Janus sequence. In the special case of no measurement collapse, this constraint correctly reproduces the expected relationship between the unitary time-evolution operators and the anti-unitary time-reversal operators \cite{sakurai2011modern}. For a single measurement, this condition may be understood as an application of quantum measurement uncollapse \cite{Korotkov2006undo,Andrew2010uncollapse,Korotkovexp2008,Kim2009,Zhong2014}. We emphasize that such an inverse operator may always be constructed as a measurement operator belonging to some POVM set \cite{Andrew2010uncollapse}. As alluded to above, there is no guarantee that the correct Janus sequences will happen; however, what is important is that such a pair of sequences is {\it physically possible}.

Switching from a forward to a backward Janus sequence generalizes the need for inverting the measurement record $r(t)$ in the qubit case of Fig~\ref{helical}. Now consider the analogous game, where a movie of the state dynamics from one of a Janus sequence of measurements is presented to us. We are not told if the sequence of measurements is $(A, B, C, \ldots)$, corresponding to the forward movie, or is instead $(\ldots, C', B', A')$, corresponding to the backward movie. We must then guess whether the movie with one of these two soundtracks is running backward or forward in time. There is no way to tell from the dynamics: Each step in each quantum state movie direction with matched soundtrack is a \emph{possible} forward evolution. 

Nevetheless, as with qubit the case before, we can still statistically discern the arrow of time. The likelihood functions to test the forward or backward hypotheses are constructed directly from the collective forward Janus measurement operator $M_F = \prod_j  M_j$, and collective backward Janus measurement operator $M_B = \prod_{j'} M_{j'}$, as used above. The probability of all of the measurement results, given (known) forward or reverse Janus sequences is $P_F(a, b, c, \ldots) = \| M_F |\Psi\ra \|^2$, or $P_B(\ldots, c', b', a') = \| M_B \Theta |\Phi \ra\|^2$, so the discriminator that generalizes Eq.~\eqref{R} is the log of their ratio $\mathcal{R} = P_F(\{j\})/P_B(\{j'\})$.

{\it Conclusions.}---We find that it is possible to time-reverse the dynamics of a quantum system, even when it is being measured.  For every nonprojective measurement, the forward measurement dynamics has an associated backwards measurement dynamics. Therefore, given a sequence of measurements and the quantum state trajectory (``the movie''), it is impossible to say whether the movie is being played forward or backward from dynamics alone. However, by examining the relative probability of whether the movie is playing forward or backward, given the measurement results (its ``soundtrack''), a statistical arrow time still emerges. We have shown how to test both aspects of the time-arrow question both in continuously measured qubits, as well as for any sequence of measurements by constructing a backwards Janus sequence which would show a possible time-reversed quantum state movie consistent with the original sequence of measurements. Our results have deep implications for fundamental questions about the time-asymmetry of the universe, as well as immediate implications for laboratory tests of quantum mechanical foundations.

{\it Acknowledgments.}---
We thank Kater Murch, Marina Cortes, and Lee Smolin for helpful discussions.
This work was supported by 
by John Templeton Foundation grant ID 58558, US Army Research Office Grants No. W911NF-15-1-0496, No. W911NF-13-1-0402, by National Science Foundation grant DMR-1506081, and by Development and Promotion of Science and Technology Talents Project Thailand.  We also acknowledge partial support by Perimeter Institute for Theoretical Physics. Research at Perimeter Institute is supported by the Government of Canada through Industry Canada and by the Province of Ontario through the Ministry of Economic Development \& Innovation.

%

\break

\appendix
\section{Supplemental Material}

In this Supplementary Material, we provide a derivation of a continuous qubit measurement that connects Eqs.~(1) of the main text to the general Janus sequence construction. We also show several examples of rare and seemingly reversed qubit trajectories. Finally, we derive the distribution of $\ln{\cal R}$ for a qubit with no Rabi drive.

\subsection{Continuous time symmetry}
For the special case of two eigenvalue observables, we can construct a Janus sequence for the diffusive continuous measurement case. To see this, consider unitary dynamics followed by partial measurement collapse, such that a density operator $\rho$ changes after a time-step $\delta t$ (up to normalization) as $\rho \rightarrow M_r U \rho U^\dagger M_r^\dagger$, where $U$ is a unitary time-evolution operator, and $M_r$ is a measurement operator indexed by a normalized result $r$ that we take to be a continuous variable. For a diffusive measurement to have a sensible continuum limit, it must come from a valid Gaussian POVM $E_r = M_r^\dagger M_r \propto \exp(-\delta t(r-A_h)^2/2\tau)$, where $A_h$ is the Hermitian observable being monitored, and $\tau$ is a characteristic measurement timescale. In this limit as $\delta t \to 0$, a succession of independent Gaussian timesteps then produces a readout $r(t)$ that is a stochastic process $r(t) = \bar{A}_h(t) + \sqrt{\tau}\,\xi(t)$, where $\bar{A}_h = {\rm Tr}[\rho A_h]$ is the moving average of $A_h$.

In the same limit, the unitary dynamics may be written to first order in time $\delta t$ as $U \approx 1 - i \delta t H/\hbar$, where $H$ is the Hamiltonian. The Gaussian POVM $E_r = M_r^\dagger M_r$ naturally factors as $M_r \propto \exp(i \delta t\, r A_{ah}/2\tau - \delta t(r-A_h)^2/4\tau)$, where we include the anti-Hermitian operator $iA_{ah}$ to allow for additional phase backaction. To first order in $\delta t$, neglecting $r^2$ as state-independent, this yields $M_r \propto 1 + \delta t(r/2\tau)A + \delta t A_h^2/4\tau$, where $A \equiv A_h + i A_{ah}$ contains Hermitian and anti-Hermitian parts. The $r$-independent term with $A_h^2$ is not reversible with any simple transformation of the record $r$; however, this term may be easily reversed by a Gaussian POVM for any observable whose square is a constant $c^2$ (implying $A_h$ has eigenvalues of only $\pm c$). As such, in what follows we will assume the form $A = {\boldsymbol \alpha} \cdot {\boldsymbol \sigma}$ of an effective qubit with Pauli matrix vector $\boldsymbol{\sigma}$, so $A_h = \text{Re}(\boldsymbol{\alpha})\cdot\boldsymbol{\sigma}$ and $A_{ah} = \text{Im}(\boldsymbol{\alpha})\cdot\boldsymbol{\sigma}$. Similarly, we assume a general qubit Hamiltonian $H = \hbar {\bf \Omega} \cdot {\boldsymbol \sigma}/2$. These considerations then lead to a (Markovian) stochastic differential equation for the normalized qubit state $\rho$ 
\be\label{eq:stoch}
\frac{d \rho}{dt} = \frac{1}{i\hbar} [H, \rho] +\frac{r}{\tau}\left[\frac{A\rho + \rho A^\dagger}{2} - {\rm Tr} \left[\frac{ A+A^\dagger}{2} \rho\right] \rho\right],
\ee
expressed in the time-symmetric (Stratonovich) picture \cite{wongzakai1965,BookOksendal} where $d\rho/dt \equiv \lim_{\delta t\to 0} [\rho(t+\delta t)-\rho(t-\delta t)]/2\delta t$. This equation reproduces Eqs. (1) in the main text if ${\bf \Omega} = \Omega {\hat y}$ and $\boldsymbol{\alpha} = {\hat z}$.

We now examine the requirements for time-reversal symmetry of Eq.~(\ref{eq:stoch}).  The time reversed solution, ${\tilde \rho}(t) = \Theta \rho(T-t) \Theta^{-1}$, must satisfy the same equation of motion (\ref{eq:stoch}).  Direct calculation indicates that is true, provided we transform to (time-reversed) operators, ${\tilde H} = \Theta H \Theta^{-1}$, and ${\tilde r}(t) {\tilde A} = - r(T-t) \Theta A \Theta^{-1}$.  This transformation is a special case of our general Janus criterion in the main text.
On physical grounds for a spin, we take the Pauli matrix vector to flip sign under time reversal, $\Theta {\boldsymbol \sigma} \Theta^{-1} = - {\boldsymbol \sigma}$, but it is straightforward to generalize this to flip the sign of only one of the Pauli matrices for a general pseudo-spin \cite{winkler2010time}.
The full inversion gives the time-reversed symmetries, ${\tilde {\mathbf \Omega}} \cdot {\tilde {\boldsymbol {\sigma}}}  = -{\mathbf \Omega} \cdot {\boldsymbol \sigma}$, and ${ \tilde  {\boldsymbol \alpha}} \cdot  { \tilde  {\boldsymbol  \sigma}} = - {\boldsymbol  \alpha}^\ast \cdot {\boldsymbol \sigma} $.  We can define time reversed quantities in one of two ways.  The first is an active transformation, which keeps the reference frame the same (${\tilde {\boldsymbol \sigma}} = {\boldsymbol \sigma}$), and the second is a passive transformation, which inverts the reference frame into a left-handed system, (${\tilde {\boldsymbol \sigma}} = -{\boldsymbol \sigma}$), thus changing the commutator structure.
The active transformation (a) dictates the mappings, ${\tilde r}_a(t) = r(T-t)$, ${\tilde {\mathbf \Omega}}_a = - {\mathbf \Omega}$ (analogous to inverting an external magnetic field),
and ${\tilde {\boldsymbol \alpha}}_a = - {\boldsymbol \alpha}^\ast$ (measuring the negated observable and reversing phase backaction), together with an inversion of the components of the Bloch coordinates, ${\tilde x}_a(t) = -x(T-t), {\tilde y}_a(t) = -y(T-t), {\tilde z}_a(t) = -z(T-t)$, which actively flips the spin.  On the other hand, the passive transformation (p) inverts the sign of the measurement readout, ${\tilde r}_p(t) = -r(T-t)$, keeps the energy definitions the same ${\tilde {\mathbf \Omega}}_p =  {\mathbf \Omega}$, and measures the same observable with reversed phase backaction, ${\tilde {\boldsymbol \alpha}}_p = {\boldsymbol \alpha}^\ast$,
while preserving the coordinates in this frame, ${\tilde x}_p(t) = x(T-t), {\tilde y}_p(t) = y(T-t), {\tilde z}_p(t) = z(T-t)$.  
With this understanding, we can see why Eqs.~(1) of the main text are invariant under the time reversal symmetry transformations discussed previously.  From the passive perspective, taking ${\tilde x}_p(t) ={x}(T-t), {\tilde y}_p(t) ={ y}(T-t), {\tilde z}_p(t) ={ z}(T-t)$ negates the left-hand side of (1) (main text), while ${\tilde \Omega}_p = \Omega$, and the reversal of the readout ${\tilde r}_p(t) = -r(T-t)$ partially inverts the right-hand side of (1) (main text).  The remaining sign reversal (effectively inverting $\Omega$) is accounted for by the sign-flipped commutation relations of the left-handed coordinate system.  From the active perspective, taking ${\tilde x}_a(t) =-{x}(T-t), {\tilde y}_a(t) =-{ y}(T-t), {\tilde z}_a(t) =-{ z}(T-t)$ keeps the time derivatives of (1) (main text) invariant, while ${\tilde \Omega}_a = -\Omega$, and ${\tilde r}_a(t) = r(T-t)$ keeps the right-hand side of (1) (main text) also invariant.

\subsection{Examples of seemingly backward-in-time trajectories}
\begin{figure}[t]
\includegraphics[width=0.7\columnwidth]{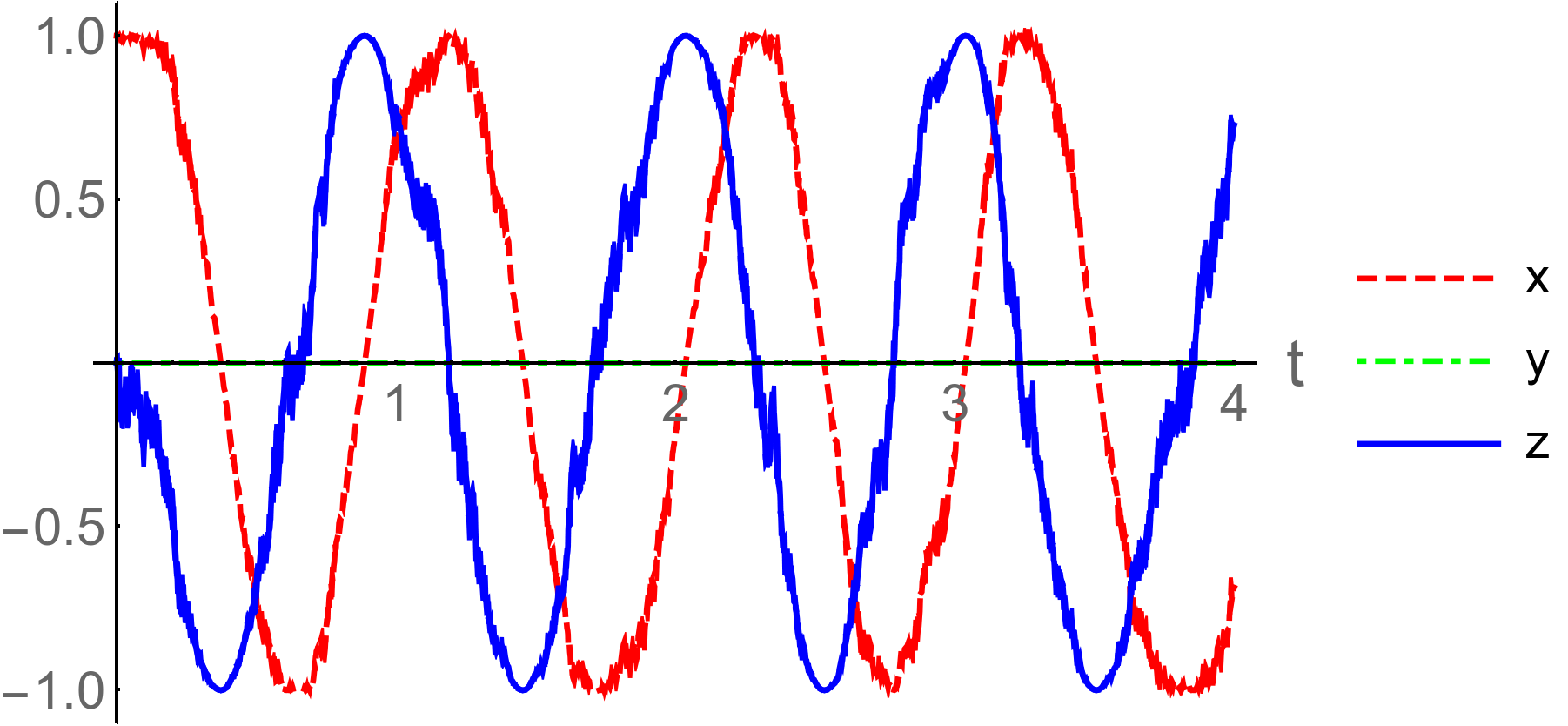}
\includegraphics[width=0.7\columnwidth]{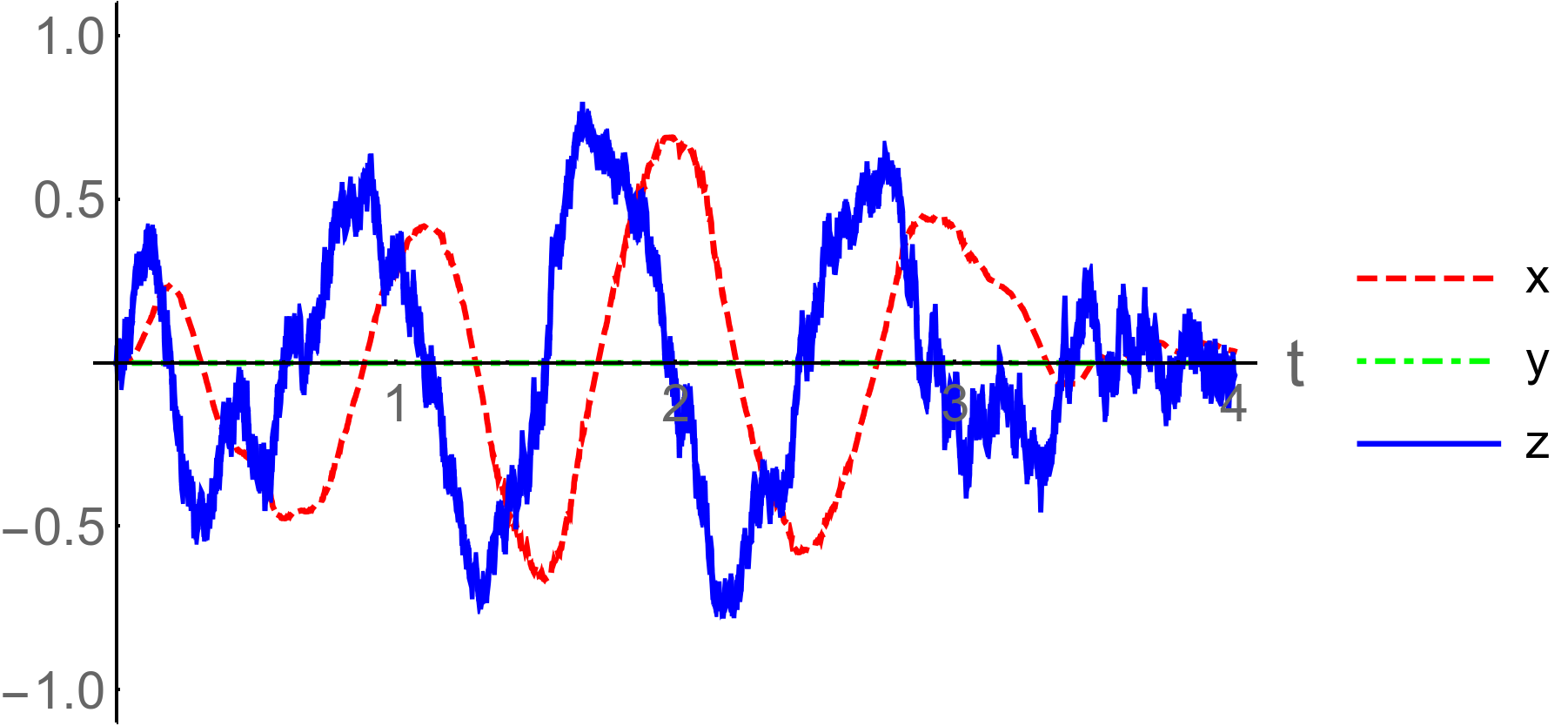}
\caption{Anomalous monitored Rabi oscillations with period $2\pi/\Omega = 0.5\tau$, measurement time $\tau = 2\,\mu$s, and duration $T = 2\tau$. (top) Pure initial state $x(t=0)=1$, with a seemingly reversed log-likelihood ratio $\ln {\cal R} = -1.40$. Comparing to the histogram in Fig.~2(d) of the main text, such a trajectory is unlikely. (bottom) Maximally mixed initial state, with a nearly symmetric log-likelihood ratio $\ln {\cal R} = 5.47\times 10^{-5}$ that is compatible with either forward or backward evolution.}
\label{backwards}
\end{figure}

As shown in Fig.~2 of the main text, for pure states undergoing monitored Rabi oscillations it is common to observe measurement runs that appear reversed (i.e., $\ln {\cal R} < 0$), even for reasonably long durations $T$. We show an example of such a seemingly reversed trajectory in Fig.~\ref{backwards} (top). If we start with a completely mixed state, however, the resulting purification of the state over time due to the measurement will reveal the directionality of time, thus seemingly preventing the evolution from being time-reversed. Nevertheless, it is still possible, though unlikely, for a trajectory to erase prior purification and return to the initially mixed state. The example in Fig.~\ref{backwards} (bottom) shows such a time-ambiguous trajectory that begins and ends with a mixed state. This example illustrates both wavefunction uncollapse as well as time-reversal invariance with no time arrow.

\subsection{Qubit with no Rabi drive}
With no Rabi drive, the forward distribution of the readout $r(t)$ consisting of $N$ independent timesteps $\delta t$ for a monitored qubit with initial z-coordinate $z^i = \text{Tr}(\sigma_z \rho^i)$ is $P_F(r) = \prod_{k=1}^N G_+(r_k) (1+z^i)/2 + \prod_{k=1}^N G_-(r_k) (1-z^i)/2$, where the Gaussian distributions $G_\pm(r_k)$ are centered at $\pm 1$, respectively, with variances $\tau/\delta t$ that define the characteristic measurement time $\tau$ for obtaining a unit signal to noise ratio \cite{Andrew2010uncollapse}. After a duration $T = \sum_{k=1}^N \delta t$ the integrated signal $\gamma = \sum_{k=1}^N r_k \delta t/T \to \int_0^T \!r(t)dt/T$ will fully determine the final qubit state. Similarly, the backwards evolution starts from the final state with z-coordinate $z^f = \text{Tr}(\sigma_z \rho^f)$ and realizes the inverted measurement sequence $-r(T-t)$ with integrated signal $-\gamma$, with the distribution $P_B(-r) = \prod_{k=1}^N G_+(-r_k)(1+z^f)/2 + \prod_{k=1}^N G_-(-r_k)(1-z^f)/2$. The arrow of time estimator $\cal R$ is thus given by
\begin{equation}
{\cal R} = \frac{P_F}{P_B} = \frac{\cosh \gamma + z^i \sinh \gamma}{\cosh \gamma - z^f \sinh \gamma},
\end{equation}
where $z^f$ is related to $z^i$ according to
\begin{equation}
z^f(\gamma) = \frac{z^i \cosh \gamma + \sinh \gamma}{\cosh \gamma + z^i \sinh \gamma}.
\end{equation}
Inserting this relation into the arrow of time estimator, after some algebra, we find the result
\begin{equation}
\ln {\cal R} = 2 \ln ( \cosh \gamma + z^i \sinh \gamma).
\end{equation}

We can directly find the probability distribution of $\ln {\cal R}$ by the relation, $P(\ln {\cal R}) d (\ln {\cal R}) = P_F(\gamma) d\gamma$. Noting the result of the derivative, $d \ln {\cal R}/d \gamma = 2 z^f(\gamma)$, we find
\begin{equation}
P(\ln {\cal R}) = \left. \frac{P_F(\gamma)}{2 |z^f(\gamma)| } \right \vert_{\gamma = \gamma(\ln {\cal R})}
\end{equation}
with an implied sum over the two solutions of $\gamma$. 

\begin{figure}[b]
\includegraphics[width=0.62\columnwidth]{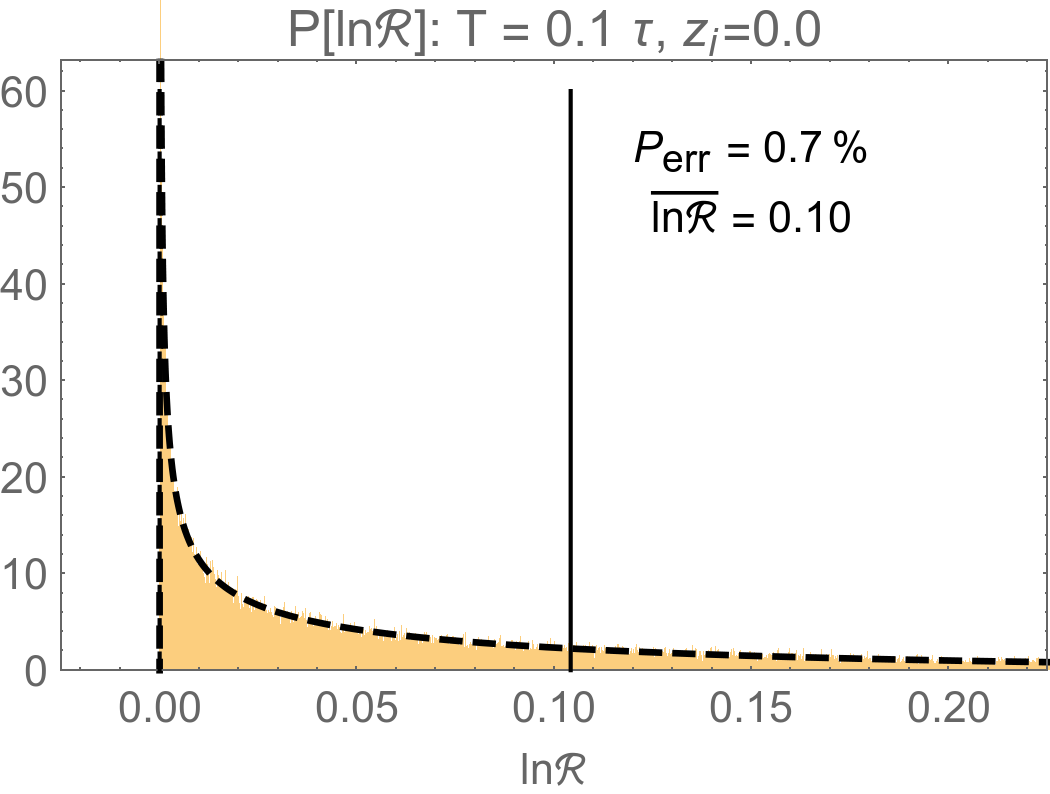}
\caption{Histogram of $\ln{\mathcal{R}}$ of $2\times 10^6$ trajectories, with $2\times 10^5$ bins, for qubit measurement with no Rabi drive from an initial state $x=1$, compared to analytics (dashed). Analytically, $P_{\rm err} = 0$; the small deviation here arises from numerical error due to the finite bin size and the divergence at $\ln {\mathcal R} = 0$.}
\label{fig:hist}
\end{figure}

The case $z^i=0$ is special because negative values of $\ln {\cal R}$ never occur. The final condition is then $z_f = \tanh \gamma$ and the solutions to the equation $x = \ln {\cal R} = 2 \ln \cosh \gamma$ are $\gamma_{\pm} = \pm \cosh^{-1} (e^{x/2})$. The distribution of $x$ thus becomes,
\begin{equation}
P(x) = \sqrt{\frac{\tau}{2\pi T}} \frac{e^x}{\sqrt{e^x-1}} \exp\left\{ -\frac{T}{2\tau} - \frac{\tau}{2T} [\cosh^{-1} (e^{x/2})]^2 \right\}.
\end{equation}
which diverges as $x^{-1/2}$ for small $x=\ln{\cal{R}}$, as shown in Fig.~\ref{fig:hist} compared to numerical simulations.

%

\end{document}